\documentclass[aps,preprint]{revtex4}%
\usepackage{amsfonts}
\usepackage{amsmath}
\usepackage{amssymb}
\usepackage{graphicx}%
\begin{document}
\preprint{ 05RBP02/SC01 (condmatsubm.tex)}
\title[Optical conductivity]{The Optical conductivity resonance from an exact description of the electronic
states around the Fermi energy.}
\author{F. Puch}
\affiliation{Departamento de F\'{\i}sica, Universidad de Zacatecas, Zacarecas, M\'{e}xico.}
\author{R. Baquero}
\affiliation{Departameto de F\'{\i}sica, CINVESTAV, A.P. 14-740, M\'{e}xico D.F.}

\begin{abstract}
In this paper we show that the optical conductivity can be calculated to agree
with experiment if the details of the electronic states around the Fermi level
are taken into account with some care. More precisely, we present a
calculation of the optical conductivity in YBa$_{2}$Cu$_{3}$O$_{7}$ on the
basis of an exact (\textit{ab initio}) three dimensional electronic band
structure calculation from which we extract the information on the bands near
the Fermi energy that can be associated to the CuO$_{2}$ plane-carrier states.
To simulate the superconducting state we superimpose a gap to these bands
alone. On these basis, we calculate from the known Kubo-Greenwood formula, the
optical conductivity in the normal and in the superconducting state. Our
calculation agrees with the experimental result even in the higher part of the
frequency spectrum. Our way of calculating the resonance suggests a model of
evolution for the bands under the effect of doping consistent with the recent
experimental findings that the optical resonance can disappear while the
sample remains superconducting. An important conclusion of this paper is that
the resonance depends mostly on the details of the electronic band structure.
It is enough to take into account the effect of the superconducting transition
through a single parameter (the gap). No details on the mechanism are needed
so no mechanism can be tested on this basis. Our calculation suggests a model
of evolution for the bands around the Fermi energy under doping that gives
some microscopic foundations to the the recent experiments that show
unambiguously that the resonance cannot be the cause of superconductivity.
Most importantly, it indicates how the background is built up and depends on
the electronic excitations accessible through values of the energy transfer on
a wider interval than those causing the resonance. These electronic
excitations determine the allowed optical transitions. From this point of
view, it is an obvious consequence that the background is with small
differences, common to all the cuprates having a CuO$_{2}$ plane. But the most
important conclusion is that the background contains essentially the same
physics as the resonance does and so does not have any detailed information on
the superconducting mechanism as well, contrary to the conclusions of recent work.

\end{abstract}
\volumeyear{year}
\volumenumber{number}
\issuenumber{number}
\eid{conductivity.tex}
\date[Date text]{27 sept. 04}
\received[Received text]{date}

\revised[Revised text]{date}

\accepted[Accepted text]{date}

\published[Published text]{date}

\startpage{1}
\endpage{ }
\maketitle

\section{Introduction.}

In a recent publication, J. Hwang, T.Timusk and G. D. Gu \cite{timusk} have
reported an infrared spectroscopy study of the optical conductivity as a
function of doping in various samples of Bi-2212. The effect of doping into
Bi-2212 is to lower both the critical temperature, T$_{c}$, and the intensity
of the resonance peak that appears in the superconducting state. This fact
represents a unique opportunity to dissociate the resonance from the mechanism
of superconductivity by means of a direct experiment. These authors have
reported the fabrication of \ several superconducting samples of Bi-2212 with
different doping content up to a particular one with 0.23 holes per Cu atom
that did not show the resonance anymore. Its high critical temperature
(Tc=55K) demonstrates that superconductivity in this sample is still robust.
This experimental fact leads to the sharp conclusion that the resonance cannot
be taken as the cause of superconductivity. Nevertheless, since the resonance
appears in the superconducting and only in the superconducting state, it
should be somehow tightly bound to the phase transition itself.

A resonance peak \cite{descubres,bourges} appears also in the spin polarized
magnetic susceptibility, $\chi_{s}(\omega)$, at a frequency, $\omega_{res}$,
that is characteristic of the specific sample. The magnetic resonance peak
appears as a common excitation to the superconducting state of all high-Tc
superconductors investigated by Inelastic Neutron Scattering (INS) so far with
a maximum T$_{c}$ $\approx$ 90K. The existence of the excitation does not
depend on the number of CuO$_{2}$ planes per unit cell: one for Tl$_{2}%
$Ba$_{2}$CuO$_{6+\delta}$, two for YBa$_{2}$Cu$_{3}$O$_{6+\delta}$ and
Bi$_{2}$Sr$_{2}$CaCu$_{2}$O$_{8+\delta}$. It has never been observed in the
monolayer system La$_{2-x}$Sr$_{x}$CuO$_{4}$ with maximum Tc $\approx$ 40K
\cite{sidis}. Although, there are several proposals in the literature
\cite{sidis} to trace the origin of the resonance to a mode of magnetic
origin, we want to point to a direct and simple relation between the effect
that the superconducting phase transition has on the electronic band
structure, and the resonance itself.

J. Hwang, T.Timusk and G. D. Gu present in their quoted paper the optical
single-particle self-energy which is directly related to the optical
conductivity. They call the resonance that they find "the resonance optical
mode". This resonance optical mode in the optical conductivity, $\sigma
_{opt}(\omega),$ is produced at the same characteristic frequency,
$\omega_{res}$, as in the susceptibility. Both resonances differ on details
though. It is important to notice that the two thermodynamic functions can be
considered mutually excluding each other in the sense that while the non-zero
contributions to the matrix elements in the spin polarized susceptibility are
intra-band, in the optical conductivity these are inter-band transitions . In
that sense it might appear at first sight surprising that the two resonances
have the same origin. Carbotte \textit{et \ al.} \cite{carbotte} assumed that
the resonance in $\chi_{s}(\omega)$ can be related to the effective spectrum
of the spin fluctuations and used it to construct the Eliashberg function of
the conventional theory of superconductivity from which all the information on
the thermodynamics follows \cite{baquerocarbotte}. They have used this
knowledge to calculate the resonance in the optical conductivity,
$\sigma_{opt}(\omega),$ in the superconducting state. Their result agrees with
experiment. They further used an inversion procedure \cite{marsiglio} that
allows to extract information about the Eliashberg function from the optical
conductivity$.$ In this way they got back their assumed Eliashberg function
(directly related to $\chi_{s}(\omega)$ in their work). The procedure used by
Carbotte \textit{et al.} establishes an essentially common origin to the
resonance in both thermodynamic functions (not to the functions themselves).
Since the experiments by J. Hwang \textit{et al.} \cite{timusk} neatly show
that the resonance in the optical conductivity is not responsible for
superconductivity, it is natural to expect that the resonance in the
susceptibility will not be responsible for superconductivity as well and
therefore cannot be directly related to the real Eliashberg function. But an
important point is that Carbotte \textit{et al.} show explicitly that the
resonances in both functions have a common origin. The resonance has been the
object of a substantial amount of work in the last years
\cite{carbotte,2enT,3enT,4enT,5enT,6enT,7enT,8enT,9enT,10enT,11enT,12enT,14enT}%
.

ARPES experiments by Lanzara \textit{et al}. \cite{lanzara} have shown that a
"kink" in the kinetic energy spectra of several cuprate superconductors
reveals a coupling of the carriers to the intermediate boson that causes the
superconducting transition. They have attributed it to phonons. Other
researchers have attributed it rather to a coupling to a magnetic mode
\cite{carbotte,15enT,16enT,17enT}. A more recent paper by Lanzara\textit{ et
al.} \cite{lanzaranuevo} emphasizes the same previous conclusion: the
intermediate boson is a phonon. It is clear \cite{timusk} that the optical
self-energy as measured by infrared is somehow related to the quasiparticle
self-energy as measured in ARPES experiments but they are not identical and
there are important differences in the two quantities \cite{carbotteprb}.

To calculate the optical conductivity in YBa$_{2}$Cu$_{3}$O$_{7}$, we start
from an \textit{ab initio} LAPW three dimensional (3D) calculation of the
electronic band structure \cite{albertotesis, albertormf, puchtesis}. When we
compared the band structure calculations in the literature among themselves,
we found that there is a certain disagreement on the exact description of the
bands around the Fermi level, E$_{F}$. To improve our results according to
experiments and to the accepted information on the bands around the Fermi
level, we found it useful to fit our 3D \textit{ab initio} bands to a
tight-binding Hamiltonian. This allows us to slightly fine-tune our bands
around the Fermi level (see below).

We, next, calculate the optical conductivity, $\sigma_{opt}(\omega),$ in the
normal state for YBa$_{2}$Cu$_{3}$O$_{7}$ from the known Kubo-Greenwood
formula \cite{f83} and compare our result to experiment \cite{puchtesis}. For
that purpose, We have identified the electronic bands of the carriers
associated to the CuO$_{2}$ plane. We then perform the same calculation in the
superconducting state. To simulate the superconducting state, we have
introduced into the electronic band structure a superconducting gap to the
bands that describe the carriers on the CuO$_{2}$ plane and only to them. We
perform the calculation using the same formula and our built up
\textquotedblright superconducting band structure\textquotedblright. The
resonance appears in the superconducting and only in the superconducting
state, at $\omega_{res}=2\Delta=38meV$ ($\Delta$ is the gap that we used for
the CuO$_{2}$-plane carrier-bands). So we argue that the superconducting phase
transition modifies effectively the electronic bands around the Fermi level
and that this feature opens several new channels for allowed transitions with
energy transfer $\omega_{res}=2\Delta$ and thus produces the experimentally
observed effect. We will show below how both the intra- and inter- band
transitions are projected by the superconducting phase transition to the same
energy $2\Delta$ a fact that explains the common origin of the resonance in
both thermodynamic functions.

The model that we present here has the advantage that, on exactly the same
footing, accounts for several experimental results of different character
\cite{baqueromodel}. We will comment on this further below. We deal in this
paper with YBa$_{2}$Cu$_{3}$O$_{7}$. Our model separates sharply the cause of
the resonance from the cause of superconductivity as the infrared experiments
\cite{timusk} indicate. The resonance arises from the effect that the
superconducting transition (the gap) has on the electronic band structure but
superconductivity (the gap itself) can arise by whatever mechanism. Further,
to the extend in which we can account for the resonance by taking only the gap
value as the information on the superconducting state, it is clear that no
information on the specific mechanism can be obtained from it. It appears that
the resonance does not contain enough information on the superconducting
mechanism to be useful to decide on it.

The background is built up from the allowed transitions at energies different
from $\omega_{res}$. The physics that it contains is essentially the same. No
information on the mechanism can be extracted from the background as well,
since it appears that the main contribution to the optical conductivity comes
from the optical allowed transitions determined by the electronic states
around the Fermi energy and by the influence that the superconducting phase
transition (solely through the gap value) has on them. The background as a
source of information on the mechanism has been suggested by Hwang et al.
\cite{timusk} solely on the basis that it is common to all HTSC. It has been
further emphasized by Norman \cite{norman}. As it appears to us, neither the
resonance nor the background in the optical conductivity can be used
effectively to decide on the mechanism of HTSC.

The situation seems not to be the same with the "kink" in the self-energy
found in the ARPES experiments \cite{lanzara}. The kink does reveal a coupling
and most probably is the key to the superconducting mechanism. Carbotte
\textit{et al. }\cite{carbotteprb} have argued that the ARPES experiments can
be interpreted as supporting either spin-fluctuations or the phonon-mediated
mechanism. They eliminate the possibility of a phonon-mediated mechanism on
the basis that their own calculation of the optical conductivity assuming this
mechanism gives a wrong dependence of this function at high frequencies. As we
shall see below, our model gives the right high-frequency dependence of the
optical conductivity. We do not assume any mechanism, we only introduce a
detailed description of the electronic bands. It is far from being clear
whether or not the phonons plays a role in HTSC. Nevertheless, in a recent
work on the isotope effect, Gweon \textit{et al. }\cite{lanzara2}\textit{
}suggest that the singlet pairing of electrons and the electron-lattice
coupling mutually enhance each other. On the basis of our work as it stands no
mechanism can be analysed. We go backwards in certain sense. We want to show
how far one can go in describing the thermodynamics of YBa$_{2}$Cu$_{3}$%
O$_{7}$ by taking the attitude that the carriers have to be described in
detail while superconductivity itself is described just by the gap value. In
conventional superconductivity the details of the phonon spectrum are crucial
while the descrition of the electrons enter just through the density of states
at the Fermi energy.We deal here with a particular experiment and with a
particular HTSC.

The rest of the paper is organized as follows. In the next section 2, we
present our electronic band structure and compare it with other work in
tliterature. \ We also include at this point our description of the
superconducting state. In section 3, we calculate the optical conductivity for
YBa$_{2}$Cu$_{3}$O$_{7}$ in the normal and in the superconducting state and
analyze our result. In section 4, we show that our model leads directly to a
suggestion for the approximate way in which the electronic band structure
might develop around the Fermi Energy under doping for the observed property
to emerge quite naturally (the resonance disappears while superconductivity
remains). In a last section 5, we put our model in perspective and draw our
conclusions. We make at this point some experimental suggestions that might
contribute to prove the usefulness of our model. In particular, we suggest
that there is no reason for the two resonances to disappear at the same doping
concentration. The "kink" in the ARPES experiments should remain as long as
the sample remains superconducting irrespective to whether it presents one,
two or no-resonance at all would it be the key experiment to reveal the mechanism.

\section{The Normal and the superconducting state of YBa$_{2}$Cu$_{3}$O$_{7}$}

\subsection{The Electronic Band Structure.}

The 3D-electronic band structure of YBa$_{2}$Cu$_{3}$O$_{7}$ has been
calculated by different \textit{ab initio }methods and by the tight-binding
method \cite{albertotesis, albertormf, f56, f65, f66, f67,  f69, f70, f71,
andersen, massida}. When we compare the different results in the literature,
we find that there are differences. The exact position of the bands can differ
in as much as 100 meV. These are important differences on the scale of meV
which is the proper scale to describe the effect of the superconducting gap.

We have first perform an \textit{ab initio} LAPW calculation of the
normal-state electronic band structure using the WIEN97 code and the
parameters of reference \cite{f71}. We found useful to fit further our
calculation to a tight-binding description as well to analyze some details of
it \cite{albertotesis. albertormf}. We have pet special attention to the
description of the band near the Fermi energy. Some features of our final
result are that we find the extended van Hove singularity around the high
symmetry point Y \cite{abrikosov} at 14 meV below the Fermi energy in
agreement with experiment \cite{campuzano}. Below the Fermi level, we also
found a van Hove singularity at about -200 meV at $\mathbf{k}=(0.42\pi
/a,0.13\pi/b,0)$ which has been reported in ref. \cite{albertormf}. The
overall features of our calculation coincide well with the rest of the other
works in the literature. We reproduce our result for the bands around the
Fermi energy in Fig. \ref{fig1}. In this figure, the bands labelled 1 and 4
belong to in-CuO-chain states while the 2 and 3 ones describe in-CuO$_{2}%
$-plane states. Notice that bands 2 and 3 that belong to the planes do not
show a significantly different dispersion from \textbf{S-X} than they do
from\textbf{ S-Y, }while bands 1 and 4 that belong to the chains do have a
different dispersion. In the upper part of Fig. \ref{fig2}, we present the
total density of states (DOS) that we get from our band structure. We have
obtained a very similar result by calculating it through the Green's function
\cite{f74} or using the tetrahedral method of integration \cite{f75, f76}. In
the lower part of the same figure, we present the DOS discriminated for each
scenario (planes, chains and c-axis). At the Fermi level, their relative
contribution is 74\% (planes), 15\% (chains) and 10\% (c-axis). The most
important contribution comes by far from the CuO$_{2}$-plane states as it is
well known.

\subsection{Description of the superconducting state.}

By whatever mechanism, the superconducting transition has the effect of
introducing a gap at the Fermi energy, E$_{F}$, on the electronic states of
the carriers affected by it. To simulate the superconducting phase transition,
we have introduced by hand a constant gap, $\Delta$, into our normal-state
electronic band structure to the bands near E$_{F}$ that can be associated to
electronic states that belong to the CuO$_{2}$ plane. It is important to say
at this point that this model allows us to reproduce right away the optical
conductivity and the spin polarized susceptibility on exactly the same basis
both in the superconducting and normal state ($\Delta=0)$. This point is
important since, as we have already recall, these thermodynamic functions seem
to exclude each other in the sense that the contributing transitions to their
corresponding matrix elements are inter-band in the first case and intra-band
in the second. We present the calculation of the spin magnetic susceptibility
in detail elsewhere \cite{baquerosus} as well as the calculation of the
tunnelling characteristics \cite{baquerotun}. Here we merely want to say that
these two results other results agree with the known experiments.

We have introduced the gap into the electronic band structure in the way that
is familiar in BCS theory \cite{schrieffer}. So, we have removed the bands
that can be associated to the CuO$_{2}$ plane from the energy interval
(E$_{F}$-$\Delta$, E$_{F}$+$\Delta$) . The states above E$_{F}$ accumulate at
the upper edge of the interval and the ones below at the bottom. Would we
introduce the gap in another way (d-symmetry, for example) the result would
differ noticeably from the experimental result. We discuss this point in
detail elsewhere \cite{baquerodwaves}. We emphasize that we introduced a gap
only to the bands formed by electronic states that can be associated with the
atoms belonging to the CuO$_{2}$ plane. In this way we obtained what we call
the "superconducting electronic band structure" for the CuO$_{2}$ plane. In
what follows we make use of our tight-binding fit to our own \textit{ab
initio} calculation to calculate the optical conductivity.

\section{Calculation of the optical conductivity.}

The dielectric function, $\epsilon=\epsilon_{1}(\omega)+i\epsilon_{2}(\omega)$
characterizes the optical properties of a material. Experimentally it can be
obtained from the reflectance spectrum. The real and imaginary part of it are
related through the Krammers-Kronig relations. The imaginary part of the
dielectric function is directly related to the optical conductivity as
$\epsilon_{2}(\omega)=\frac{4\pi\sigma_{opt}(\omega)}{\omega}.$ For inter-band
transitions, we can calculate the optical conductivity from the
\textit{Kubo-Greenwood} formula \cite{f83}%

\begin{equation}
\sigma_{\text{%
\symbol{188}%
}}(\omega)=-\frac{\pi e^{2}h^{2}}{m^{2}\omega\Omega}\sum_{l,n}\int dkP_{\ln
}^{i}P_{nl}^{j}f_{n}(\mathbf{k)[}1-f_{l}(\mathbf{k)]}\delta(E_{l}%
(\mathbf{k})-E_{n}(\mathbf{k)-}h\omega) \label{kubof}%
\end{equation}

where%

\begin{equation}
P_{\ln}^{i}=\text{ }<\Psi_{l}(\mathbf{k})|\frac{\partial}{\partial x_{i}}%
|\Psi_{n}(\mathbf{k})> \label{optical}%
\end{equation}
is the optical transition matrix. $|\Psi_{n}(\mathbf{k})>$ is the Bloch
function for the n-band and \textbf{k} is the wave vector defined in the first
Brillouin zone. $E_{n}(\mathbf{k)}$ is the corresponding band energy,
$f_{n}(\mathbf{k)}$ is the Fermi-Dirac distribution function, $\omega$ is the
frequency of the radiation and $\Omega$ is the volume of the unit cell. The
rest are known constants. Now we expand the Bloch function in term of orbital
functions as%

\begin{equation}
|\Psi_{n}(\mathbf{k})>=\frac{1}{\sqrt{N}}\sum_{\alpha,j}u_{n,\alpha
}e^{i\mathbf{k.r}_{j}}|\varphi_{\alpha}(\mathbf{r-r}_{j})>. \label{bloch1}%
\end{equation}

where N is the number of unit cells, $\varphi_{\alpha}(\mathbf{r-r}_{j})$ is
the orbital wave function with quantum numbers $\alpha$ and $\mathbf{r}_{j}$
is the origin of the j-th unit cell. The $u_{n,\alpha}$ are the coefficients
of the expansion. Substituting Eq. \ref{bloch1} into Eq. \ref{optical}, we get%

\begin{equation}
P_{\ln}^{i}=\sum_{\alpha,\beta}u_{l,\alpha}u_{n,\beta}\sum_{m,j}%
M_{ijm}^{\alpha,\beta}e^{i\mathbf{k.(r}_{j}\mathbf{-r}_{m})} \label{optical1}%
\end{equation}

with%

\begin{equation}
M_{ijm}^{\alpha,\beta}=\text{ }<\varphi_{\alpha}(\mathbf{r-r}_{m}%
)|\frac{\partial}{\partial x_{i}}|\varphi_{\beta}(\mathbf{r-r}_{j})>
\label{elemento}%
\end{equation}
The matrix element, Eq. \ref{elemento}, is zero except for in-site and first
nearest neighbors transitions. $M_{ijm}^{\alpha,\beta}$ can be calculated from
our tight-binding fit if we use an approximation suggested by Harrison
\cite{f84}. Within this approximation, for inter-site transitions (atom at
$\mathbf{r}_{j}$ with atom at $\mathbf{r}_{i}$), $M_{ijm}^{\alpha,\beta}$ can
be clast proportional to $\frac{x_{m}}{d^{2}}$, where $x_{m}$ is the
m-component of the vector $\mathbf{r}_{j}-\mathbf{r}_{i}$ and $d=|\mathbf{r}%
_{j}-\mathbf{r}_{i}|$. For intra-site transitions due to symmetry
considerations, $M_{ijm}^{\alpha,\beta}$ is zero whenever the difference
between the angular momentum projections $l_{\alpha}-l_{\beta}$ is even and
different from zero otherwise.

\subsection{The normal state}

We have calculated Eq. \ref{kubof} in 1/8 of the first Brillouin zone (FBZ) at
$T=0K$ therefore the Fermi functions were put equal to 1 for energies below or
equal to the Fermi energy, $E_{F}$, and 0 otherwise. We show in the next Fig.
\ref{fig3} three components of the optical conductivity tensor (Eq.
\ref{kubof}) namely $\sigma_{xx}$, $\sigma_{yy}$, and $\sigma_{zz}$ so that we
can compare our results with the ones in the literature. Fig. \ref{fig3} (top)
shows our results for an energy interval from 1-10 eV. Fig \ref{fig3} (bottom)
gives low-energy details (0-0.5 eV) of the three tensors. There is a clear
anisotropy at low energies between $\sigma_{xx}$ and $\sigma_{yy}$. This is
due to transitions that take place on the chains ($y$-direction) that
contribute to $\sigma_{yy}$ but not to $\sigma_{xx}$. Garriga \textit{et al.
}\cite{f85} report measurements on $\varepsilon_{2}(\omega)$ which show a
maximum around $\omega=8eV$ , a peak at $4-5eV$ and a minimum around $2-3eV$.
Tajima \textit{et al. }\cite{f88} report similar results at high energies and
a minimum around $6eV$. Our calculation agrees well with these experimental
results and with theoretical calculations at low energy \ref{f89} and at
medium and high energies \cite{f86,f87}.

\subsection{The superconducting state}

We have calculated the optical conductivity from Eq. \ref{kubof} in 1/8 of the
FBZ with 64 points per axis. We have used our electronic band structure where
a gap was inserted in the way described above. We allow only inter-band
transitions on the in-CuO$_{2}$-plane states. Our result for $\sigma(\omega
)$\ in the normal and in the superconducting state appears in the next Fig.
\ref{fig4}. As we can see in this figure, the effect of the transition is to
shift the spectral weight of the almost featureless $\ \sigma(\omega)$
function to higher energies, namely above 30 meV, producing the sharp
resonance at 38 meV. This occurs because all the allowed transitions within
bands (one below and one above the Fermi level) that differ in the normal
state by less than $2\Delta$ are projected in the superconducting state to the
gap edges (above and below E$_{F}$) and the transition will take place at
$2\Delta$ irrespective to the energy at which it occurs in the normal state.
We have used $\Delta_{plane}$=19 meV for YBa$_{2}$Cu$_{3}$O$_{7}$. Both, the
calculation in the normal state and the one in the superconducting state,
reproduce the experimental results. Our model does not assume any mechanism
whatsoever. Therefore an interesting point that arises is that on quite
general grounds, namely that the phase transition introduces a gap to the
electronic band structure, the resonance can be reproduced. This fact casts
some doubts on whether at all the optical conductivity has enough information
to allow a sharp decision on the mechanism (neither in the resonance nor in
the background).

\section{Superconductivity with and without the resonance}

A suggestion based on our model on how the effect of doping can cause the
total disappearance of the resonance while the sample remains superconducting
is sketched on Fig. 5. Let the band (a) contains a possible initial state (i)
below the Fermi energy, E$_{F}$ (dash-dot line). The energy difference between
state (i) and E$_{F}$ let it be less than the gap associated to the plane,
$\Delta_{plane}$ (dashed line). As the contributing transitions are
inter-band, the final state has to lie on a different band (b) above the Fermi
level. The transitions are direct (no momentum transfer). In the normal state
this transition contributes to the resonance at a frequency equal to the
energy difference between the final and the initial state. In the
superconducting state, this and several similar events will contribute as well
but at an energy 2$\Delta$ due to the effect that the superconducting
transition has on the bands. It is enough that, upon doping, the bands evolve
so that their energy difference in the normal state gets higher than 2$\Delta$
at this particular point of the FBZ where the transition occurs for this
particular event to cease to contribute to the resonance. It will contribute
at a higher energy and, consequently, the resonance will have one less event
that contributes and its spectral weight diminishes. Eventually the resonance
disappears. It is obvious, on the other hand, that bands at certain doping
that did not contribute at a previous one can contribute, but the net effect
can be a lost of contributing events due to doping. The exact issue depends on
the details of the bands and on the specific influence of the doping on them
and on the value of the gap for each family of compounds and it has to be
calculated in detail. We suggest that this is what happens in the Bi-family
\cite{timusk} where experiment shows the disappearance of the resonance with
doping. Under doping, the gap shrinks at a slower pace than the number of
events contributing to the resonance and, consequently, the resonance
disappears but superconductivity remains.

\section{Conclusions}

We have shown in this paper that the resonance in the optical conductivity
that appears in the superconducting and only in the superconducting state can
be obtained from an ab initio three dimensional calculation just by
introducing to the electronic bands that can be associated to the CuO$_{2}$
plane a gap, $\Delta$. The actual calculation of the optical conductivity is
two-dimensional (in-plane). The model produces the resonance at 2$\Delta$ and
therefore we have consequently introduced into the calculation $\Delta
=19meV$\ to reproduce the resonance at the right experimental frequency. The
curves agree very well with experiment in the normal and the superconducting
state. It is interesting that we get the experimental trend of the function at
high frequency both in the normal and in the superconducting state. We do not
assume any mechanism in our calculation. We have further suggested a way in
which this model can account for the experimental results on the effect of
doping on the intensity of the resonance \cite{timusk}.

It is important to say that using exactly the same model, we have calculate
the spin polarized susceptibility in the normal and in the superconducting
state and reproduce the experimental results. We obtain that this resonance
could disappear as well but that there is no reason for it to disappear at the
same doping level than the one in the optical conductivity \cite{baquerosus}.
The model does not reproduce the experimental results for the tunneling
experiments unless it is extended to three dimensions. If we impose further to
the 3D bands obtained from our ab initio calculation, an additional gap to the
states that can be associated to the chains ($\Delta_{chians}=7meV)$ and keep
$\Delta=0$ for the bands that can be associated to the c-axis, we reproduce
the tunnelling characteristics in agreement with experiment \cite{baquerotun}.
A somehow similar approach in the sense that they used a different value for
the gap on each scenario (planes, chains and c-axis) has been used before
\cite{cucolo} to simulate successfully experimental results on tunnelling,
specific heat and ultrasonic attenuation. On that basis we expect to reproduce
these two last results as well from our more detailed model.

A more delicate point is to reproduce the temperature behavior of the
resonance. The resonance frequency hardly changes in the range from zero to
the superconducting temperature but, on the contrary, its intensity is very
sensitive to it, decreases with increasing temperature and vanishes steeply at
T$_{c}$ \cite{sidis}. We will suggest that this is a combined effect of the
behavior of the superconducting gap with temperature and the separate effect
of the increasing temperature on the electronic band structure.itself. The
effect of temperature on the electronic bands itself has never been considered
before in this context but within the critical temperature range ($\approx$100
K $\approx$ 10 meV) also the temperature itself could have a non-negligible
effect on the electronic band structure at the meV scale.\ Calculations of
this effect in metals like Cu and Ni have been made in the past by Delgadillo
\textit{et al. }\cite{delgadillo}. They find that the electronic bands around
the Fermi level do displace themselves as an effect of temperature in the meV
scale.This fact and the known temperature dependence of the superconducting
gap itself might explain (always within the same model) the behavior of the
resonance with temperature. Notice that 10 meV, is of the order of $\Delta/2$,
where $\Delta$ is the superconducting gap associated to the CuO$_{2}$ plane
\cite{baquerotemp}.

The goal of the work presented in this and some other papers is to show how
far one can go in describing the experimental results starting from what could
be said "the result" of a theory of superconductivity for high-Tc
superconductors. One possible conclusion is that the detailed description of
the carriers plays an important role in HTSC a fact that is in sharp contrast
with conventional superconductivity where the electronic band structure
characteristics enter merely through the density of states at the Fermi
energy. as we atated above.

\bigskip\newpage

FIGURE CAPTIONS

Fig.1 3D-electronic band structure of \ YBa$_{2}$Cu$_{3}$O$_{7}$ near the
Fermi level. Bands 2 and 3 are in-CuO$_{2}$-plane states; bands 1 and 4 belong
to in-CuO-chain states. Notice that the dispersion in the interval S-X is not
very different from the one in the interval S-Y for the bands labbelled 2 and
3 (the in-CuO$_{2}$-plane states). This is not true for bands labbelled 1 and
4 that are in-CuO-chain states. The X-Y symmetry is not expected on this scenario.

\bigskip

Fig.2  The total density of states (a) and the contribution of each scenario (b).

\bigskip

Fig.3  In the upper part (a), we show the optical conductivity-tensor
components $\sigma_{xx}$, $\sigma_{yy}$, and $\sigma_{zz}$ in the normal state
in a scale 0-10 meV. In the lower part (b), we show the same functions in more
detail below 0.5 eV so that our results can be compared easily with the ones
in the literature (see text).

\bigskip

Fig. 4 The optical conductivity function, $\sigma(\omega)$, in the normal
(full line) and in the superconducting state (dots). Notice that at higher
energies, $\sigma(\omega)$, increases with energy in the superconducting state.

\bigskip

Fig. 5 The dash-dot line is the Fermi energy and the dash-lines above and
below it are the gap edge in the superconducting state. Let us consider
allowed transitions in the normal state. In the band (a) we have selected a
possible initial state (i) and on the band (b) a possible final state. In the
normal state, this tansition contributes to the resonance at a frequency equal
to the energy difference between (f) and (i). In the superconducting state,
both (f) and (i) will be projected to the gap edge so that this and several
other low-lying similar transitions will contribute all at the same energy
2$\Delta$ due to the effect of the superconducting transition on the bands.
Upon doping, it is enough that the bands evolve so that their energy
difference in the normal state gets higher than 2$\Delta$ at this pont of the
FBZ (dot-line) for this particular event to cease to contribute to the
resonance (see text).

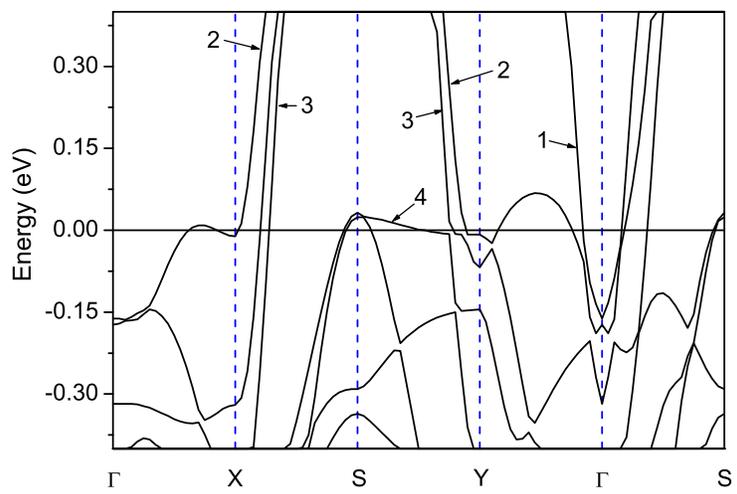
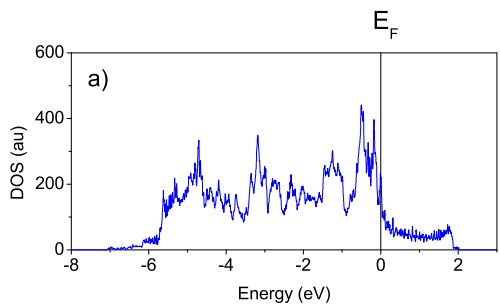
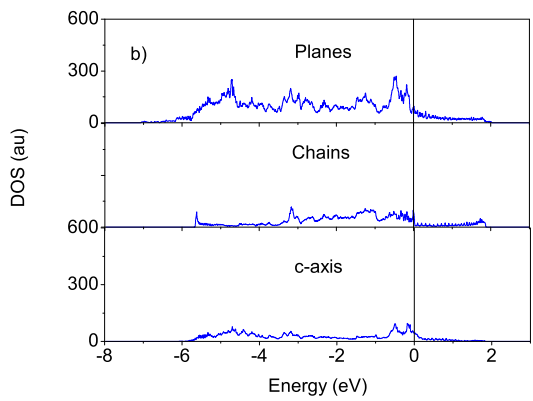
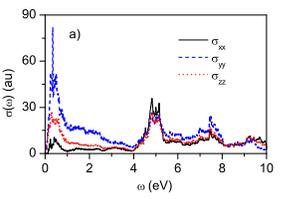
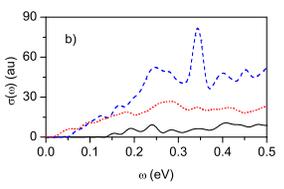
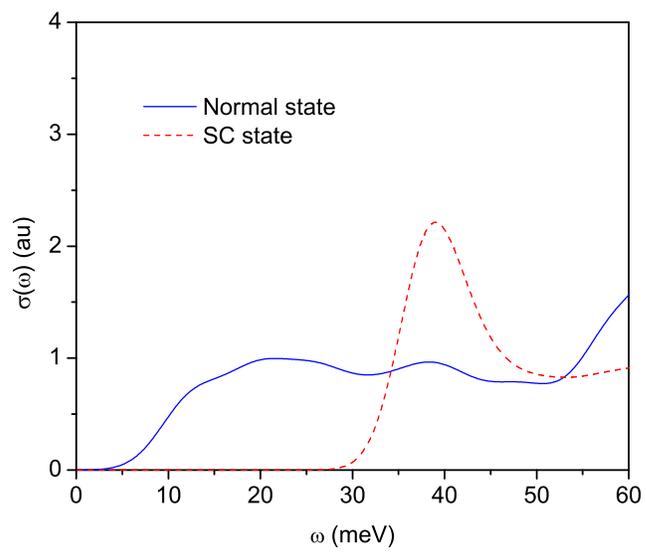
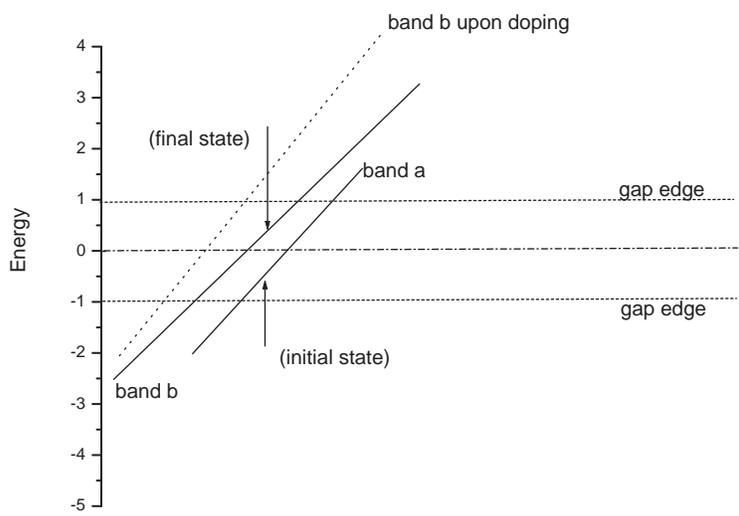
\end{document}